\documentclass[a4paper,11pt]{article}

\usepackage[a4paper, total={6in, 8in}]{geometry}
\usepackage{amsmath,amssymb,amsthm,amscd,verbatim,graphicx}
\usepackage[mathscr]{eucal}
\usepackage{bbm} 
\usepackage{enumerate} 
\usepackage{enumitem} 
\usepackage{floatrow} 
\usepackage{tikz}
\usepackage{float} 
\usepackage{cprotect} 
\usepackage{caption}
\usepackage[framemethod=tikz]{mdframed} 
\usepackage{xparse, adforn} 
\usepackage{listings} 
\usepackage{color} 
\usepackage{bm} 
\usepackage{xcolor} 
\usepackage{hyperref}
\usepackage{authblk} 
\usepackage{cleveref} 
\usepackage{array}
\usepackage{graphicx}
\usepackage{multirow}
\usepackage{floatrow}
\usepackage{apacite} 
\usepackage{natbib}
\usepackage{dirtytalk}
\usepackage{epstopdf}
\AppendGraphicsExtensions{.tif}

\hypersetup{ 
    colorlinks=true, 
    linkcolor=blue,
    filecolor=magenta,      
    urlcolor=blue,
    citecolor=blue
}

\usepackage{listings}
\usepackage{xcolor}
\definecolor{dkgreen}{rgb}{0,0.6,0}
\definecolor{gray}{rgb}{0.5,0.5,0.5}
\definecolor{mauve}{rgb}{0.58,0,0.82}
\makeatletter
\lstdefinestyle{mystyle}{
  basicstyle=%
    \ttfamily
    \color{blue}%
    \lst@ifdisplaystyle\scriptsize\fi
}
\makeatother

\makeatletter
\renewcommand\verbatim@font{\color{green!30!black}\normalfont\ttfamily}
\makeatletter

\newcommand{\deq}{\mathbin{\raise0.4pt\hbox{:}{=}}}

\title{SARGDV: Efficient identification of groundwater-dependent vegetation using synthetic aperture radar}

\author[1]{Mason Terrett\thanks{M.Terrett@latrobe.edu.au}}
\author[3]{Daniel Fryer\thanks{Daniel.fryer@uq.edu.au}}
\author[2]{Tanya Doody\thanks{Tanya.Doody@csiro.au}}
\author[1]{Hien Nguyen\thanks{H.Nguyen5@latrobe.edu.au}}
\author[2]{Pascal Castellazzi\thanks{Pascal.Castellazzi@csiro.au}}
\affil[1]{Department of Mathematics and Statistics,
La Trobe University, Bundoora VIC 3083, Australia}
\affil[2]{Commonwealth Science and Industrial Research Organisation (CSIRO),
Land and Water, Waite Rd, Urrbrae SA 5064, Australia}
\affil[3]{School of Mathematics and Physics, 
The University of Queensland,
St Lucia QLD 4067, Australia}

\date{August 2020}

\providecommand{\keywords}[1]{\textbf{\textit{Keywords---}} #1}

\begin{document}
\maketitle

\section{Abstract}

Groundwater depletion impacts the sustainability of numerous groundwater-dependent vegetation (GDV) globally, placing significant stress on their capacity to provide environmental and ecological support for flora, fauna, and anthropic benefits. Industries such as mining, agriculture, and plantations are heavily reliant on groundwater, the over-exploitation of which risks impacting groundwater regimes, quality, and accessibility for nearby GDVs. Cost effective methods of GDV identification will enable strategic protection of these critical ecological systems, through improved and sustainable groundwater management by communities and industry. Recent application of synthetic aperture radar (SAR) earth observation data in Australia has demonstrated the utility of radar for identifying terrestrial groundwater-dependent ecosystems at scale. This provides a foundation for the development of improved satellite imagery-based classification models that overcome the limitations of typical multispectral earth observation data. We propose a robust classification method to advance identification of GDVs at scale using processed SAR data products adapted from a recent previous method. The method includes the development of SARGDV, a binary classification model, which uses the extreme gradient boosting (XGBoost) algorithm in conjunction with three data cubes composed of Sentinel-1 SAR interferometric wide images. The images were collected as a one-year time series over Mount Gambier, a region in South Australia, known to support GDVs. The SARGDV model demonstrated high performance for classifying GDVs with 77\% precision, 76\% true positive rate and 96\% accuracy. This method may be used to support the protection of GDV communities globally by providing a long term, cost-effective solution to identify GDVs over variable regions and climates, via the use of freely available, high-resolution, globally available Sentinel-1 SAR data sets. Our method offers global water management agencies a means toward more sustainable management of regional groundwater resources by providing an efficient method to identify significant GDV occurrence within areas where substantial groundwater extraction is ongoing.

\keywords{groundwater-dependent ecosystems,
    groundwater,
    GDE,
    vegetation,
    water resource management,
    remote sensing,
    Sentinel-1,
    wetland,
    boosted tree learner}
\section{Introduction}

Groundwater serves as a highly valuable natural resource for both human communities and the natural environment, offering a stable, uncontaminated source of water for consumption, agricultural use, and numerous industries. The utility of groundwater for  continued development, sustainability and advancement of global communities is well documented in its significance \citep{World.2017, Smith.etal.2016}. For industry, in particular, there is heavy reliance on groundwater for mining and irrigation \citep{Doody.2017}, which \cite{Siebert.2010} denotes as the most important sector for global water use, responsible for approximately 70\% of the total fresh water and 90\% of drinkable water extraction and use.

In terms of human consumption, globally, many communities and nations rely on groundwater as their primary source of fresh drinking water. This is mainly due to groundwater being a relatively stable and safe resource that is largely free from pathogenic risks, which can impact the public health of communities. Surface water sources on the other hand, can become readily contaminated through human and animal interaction \citep{World.2017}. In addition to mitigating contaminants and microbial concerns, groundwater is also considered to be a safer alternative to surface water in terms of general chemical and radiological risks \citep{World.2017}.

Numerous reports indicate that the consequences of groundwater resource depletion is expected to severely impact many countries and global communities in a variety of ways, presenting significant challenges to the health, social, environmental and economical sustainability of these communities \citep{Aslam.et.al.2018, Khare.Varade.2018, World.2017}. Several large aquifers are seeing significant levels of over exploitation through irrigation in countries such as Mexico, China, Iran, USA, Saudi Arabia, Pakistan and India. The ongoing over extraction of groundwater resources in several of these regions is surpassing the recharge rate of groundwater systems which is expected to impact natural groundwater discharge \citep{Jakeman.2016, wadaGlobalDepletionGroundwater2010}.

Continuous over-extraction of groundwater above sustainable extraction limits, places significant stress on existing vegetation and ecological systems which rely on groundwater resources to ensure long-term survival. Such is the case for groundwater-dependent ecosystems (GDEs), which serve as ecologically significant support structures for flora, fauna, and human communities globally \citep{Doody.2017, Van.2018, Aslam.et.al.2018}.

\cite{Richardson.2011} define GDEs as \say{natural ecosystems that require access to groundwater to meet all or some of their water requirements on a permanent or intermittent basis so as to maintain their communities of plants and animals, ecological processes and ecosystem services}. There are three main categories of GDE: \emph{subterranean} (aquifers and cave ecosystems), \emph{aquatic} (ecosystems which are dependant on surface expression of groundwater, such as springs, swamps and river baseflow systems), and \emph{terrestrial} (ecosystems dependent on the subsurface presence of groundwater, such as vegetation complexes, and some riparian vegetation communities) \citep{Doody.etal.2019}.

Terrestrial GDEs, or groundwater-dependent vegetation (GDVs), serve as a crucial resource for several species of flora and fauna and provide numerous critical ecological services, which include the decontamination of water resources; maintenance of carbon/nitrogen cycle; food, habitat and biodiversity corridors for local faunal migration, as well as numerous additional benefits for the local biosphere \citep{Dresel.2010, Gow.2016, Perez.2016}. A decline in GDV health can therefore impact local flora, fauna, and human communities, the latter of which rely on GDVs to assist with providing clean water in addition to social and economic benefits \citep{Doody.2017, Aslam.et.al.2018, Perez.2016}.

Due to their reliance on groundwater, depletion or contamination of subsurface water can incur accumulated stresses, which directly impact the condition and subsequent survival of GDVs. Over extraction of groundwater near GDVs can destabilise potential water transmissivity during critical periods in their life cycle. Likewise, the continued onset of environmental stresses as a result of climate change is expected to affect the capacity for surface/groundwater interactions, influencing the temporal behaviour and reducing the ability of GDVs to endure locally, regionally and globally \citep{Doll.2009, Dresel.2010, Klove.2014}.

In order to protect GDVs, it is necessary to adequately distinguish between vegetation that are not reliant on groundwater and those that are. Previous research has shown this to be a challenging task as groundwater extraction by vegetation species is difficult to quantify without field-based observation methods, such as sap flow, isotopes studies, and the use of bore hole instrumentation \citep{Doody.etal.2019}.
Despite being the most accurate, field-based methods typically require substantial time, effort and resources, and in turn only provide local evaluation in locations where GDVs can span a substantial area \citep{Doody.2017}.

Identification and mapping of GDVs at scale has been an ongoing scientific endeavour for over a decade, with numerous research efforts implementing robust spatial technologies such as Multispectral Satellite (MS) imagery to classify GDVs \citep{Eamus.Froend.2006, Dresel.etal.2010, Barron.et.al.2014, Eamus.2015}. While application of remote sensing technologies have shown promise, continued research in this space is required to improve and extend upon previous methods, where data gaps are present (i.e. tropical cloudy regions) \citep{Doody.2017, Castellazzi.et.al.2019}.

To overcome the limitations of the previously established methods, \cite{Castellazzi.et.al.2019} presented a novel method to classify GDVs using Synthetic Aperture Radar (SAR) data, a remote sensing system which employs the use of microwaves for earth observation \citep{chan.2008}. Similar to MS systems, SAR offers large scale earth observation imagery, however, while it does not provide image data with Red-Green-Blue colour channels, SAR offers additional benefits over conventionally used MS systems.

SAR systems, such as the European Space Agency’s Sentinel-1 project, are cloud and weather invariant, and self-illuminating (i.e. they do not depend upon reflected sunlight for earth observation) \citep{chan.2008, Franceschetti.Lanari.1999}. Such properties allow SAR to consistently collect visibly clear, high quality, and high-resolution terrain data at all times, irrespective of weather conditions. MS systems are limited such that earth observation data can only be collected when reflected sunlight is available, and additionally, inclement weather and cloud cover prevent collection of terrain information.

In Australia, \cite{Castellazzi.et.al.2019} incorporated the use of Sentinel-1A SAR Interferomic Wide data and time series analysis over two Australian regions --- temperate Mount Gambier in South Australia and the Wildman-Kakadu area of a tropical floodplain region in the Northern Territory. The method involved the development of a prototype GDV index, referred to as SARGDE, which uses sequential time series SAR data to classify GDEs. SARGDE provides a useful, and repeatable technique, demonstrating high accuracy at classifying GDVs on a pixel-by-pixel basis. A combination of three SAR data cubes, Vertical-Horizontal (VH), and Vertical-Vertical (VV) SAR intensity images, and InSAR coherence images (CC), were used. The data cubes were derived from the SAR C-band wavelength, and are considered to contain information which correlates to known vegetation signal interaction data characteristic of GDVs (see \citealt{Castellazzi.et.al.2019}).

While this method has shown promising results for the chosen study regions, deriving the SARGDE index from SAR imagery relied heavily on assumptions of InSAR coherence and SAR intensity values based on expected GDV structural characteristics.
Specifically these include: 1) low InSAR coherence of dense vegetation, 2) limited seasonal changes of InSAR coherence, and 3) relatively limited seasonal change in VH intensity values. Lastly, cross-validation, which is useful for assessing the model effectiveness and to prevent overfitting, was not included in the model fitting procedure \citep{Castellazzi.et.al.2019}.

A complementary direction of inquiry includes removing prior knowledge from the model, and instead allowing it to learn algorithmically from observed data. This assumption-free approach gives the model more freedom to capture a larger amount of observed information. Additionally, once the model has learned from the available data, it can be interrogated to provide insights into the relationship between SAR images and GDV attributes.
In light of this, an improved method which extends upon the learnings and overcomes the limitations from \cite{Castellazzi.et.al.2019} to deliver high accuracy classification of GDVs over a region of interest (ROI) is viable to pursue.

Thus, we propose a novel method for the development of a GDV classification model, referred to as SARGDV, which uses a machine learning algorithm in conjunction with similar data cubes adapted from \cite{Castellazzi.et.al.2019}.
The model is designed to be generalisable, adapting well to new, previously unseen data, drawn from the same distribution used to create the model, and uses the contemporary XGBoost algorithm, as it is well regarded in its use for classification tasks, and is versatile, such that it makes less assumptions about the data set, and is able to fit a wide range of data scenarios \citep{Chen.Guestrin.2016}.
\cite{Fryer.etal.2020a} provides recommendations for improvements to SARGDE, including the addition of spatial smoothing via markov random fields. In adapting this recommendation, conditional random field (CRF) smoothing \citep{Plath.etal.2009} is implemented, to improve prediction accuracy by accounting for spatial correlation. Cross-validation is employed, and standard techniques such as receiver operating characteristic (ROC), and precision-recall curve (PRC) plots are used to assess model performance. Additionally, a logistic regression benchmark, and depth-to-water interpolation map are also used to evaluate and validate SARGDV.

The objective of this study is to build upon the learnings from \cite{Castellazzi.et.al.2019} and provide an economical, reproducible method for the development of a robust and machine learning classifier for GDVs. We expect that our method can serve as a useful tool, which may provide institutions, such as government and industry bodies, a long-term solution for identification, management, and preservation of GDVs. Additionally, this may support institutions to better manage their GDV inventories through improved water regional resource management, and designation of protected vegetation zones to mitigate potential impact by industry. The insights provided with respect to location of GDVs, can instigate additional, targeted hydrological studies that further our understanding of timing and duration of vegetation connection to groundwater and improve understanding of the hydrological processes that support GDV persistence. 

Lastly, the method reported, is the first time a boosted tree algorithm for identifying GDVs using SAR data has been developed. The novelty of integrating domains of machine learning, environmental monitoring, and SAR data products, enables additional research efforts to be pursued at the intersection of these fields.

\section{Methods}

\subsection{Study location}
\begin{figure}[!h]
    \centering
    \includegraphics[scale = 2.0]{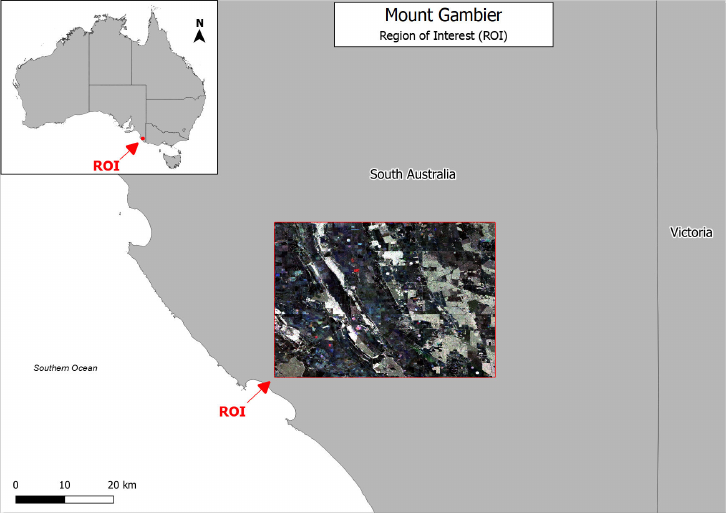}
    \caption{The chosen ROI within the Mount Gambier region of South Australia with overlay of Sentinel1 SAR image example.}
    \label{fig:ROI}
\end{figure}

For our study ROI, we have chosen a subset of the Mount Gambier (37.8$^{\circ}$S, 140.8$^{\circ}$E) from \cite{Castellazzi.et.al.2019}, located in the South-East of South Australia, comprising an area of $\approx$1800$\text{km}^2$ (figure \ref{fig:ROI}).

The ROI is located in a region known as the “Green Triangle”, which covers an area of approximately $7{,}000{,}000$ hectares \citep{Clancy.2011}. 
The climate is considered mediterranean, with dry, warm to hot summers, and cold, moist winters \citep{benyonImpactsTreePlantations2006}. The annual average rainfall of Mount Gambier in 2017, 2018 (the acquisition periods of data cube imagery) were 712.4mm, and 713.1mm, respectively, with both periods recording slightly above the annual average rainfall (114\%, and 108\%, respectively) \citep{climate.2020}.
Most rainfall does not evapotranspire; instead draining vertically into unconstrained aquifers to recharge them \citep{benyonImpactsTreePlantations2006}.

The region is host to numerous plantation-based industries, with hardwood and softwood plantations comprising $178{,}000$ and $170{,}000$ hectares, respectively. The main species grown include blue gum (\emph{Eucalyptus globulus}) and radiata pine (\emph{Pinus radiata}), with little commercial wood production of native species \citep{Clancy.2011}. The plantations were grown over areas with shallow water-tables of which the vast majority of available groundwater exists in permeable, unconstrained or semi-constrained aquifers generally ${<}$20m underground, with some areas only a few metres underground \citep{benyonImpactsTreePlantations2006}. 

The presence of GDVs has been established from previous research conducted by \cite{benyonImpactsTreePlantations2006}, \cite{Castellazzi.et.al.2019} and \cite{Wood.Harrington.2015}. In choosing this ROI, it is possible to make direct comparisons with results from \cite{Castellazzi.et.al.2019} when exploring potential improvements to the SARGDE method. Additionally, using the Water Connect groundwater data portal (\url{https://www.waterconnect.sa.gov.au/Systems/GD/Pages/Default.aspx}), borehole observations within the ROI have been acquired which provides measurements of observed depth to (ground) water (DTW). Both \cite{Doody.etal.2019} 
and \cite{SKM.GDE.Atlas.2012} outline DTW as a significant determinant for likelihood of groundwater dependence of vegetation species, with shallower DTW providing a higher likelihood of overlying vegetation being GDVs \cite{Doody.2017}. Thus these observations provide an additional source of validation of the hydrogeological activity within the ROI.

\subsection{Development tools}
Two robust open source software tools were used to analyse and process the SAR data cubes used in the development of the SARGDV model. The R programming language \citep{R.2020} via Rstudio is employed to analyze and process SAR data, and build the model via a custom R package produced by the authors (\url{https://github.com/frycast/rsar}).
QGIS 3.8, an open source geographic information system software toolbox for spatial data, is used to visualise and analyze various raster and vector data sets.

\subsection{Model data sets}
\begin{figure}[!h]
    \centering
    \includegraphics[scale = 2.1]{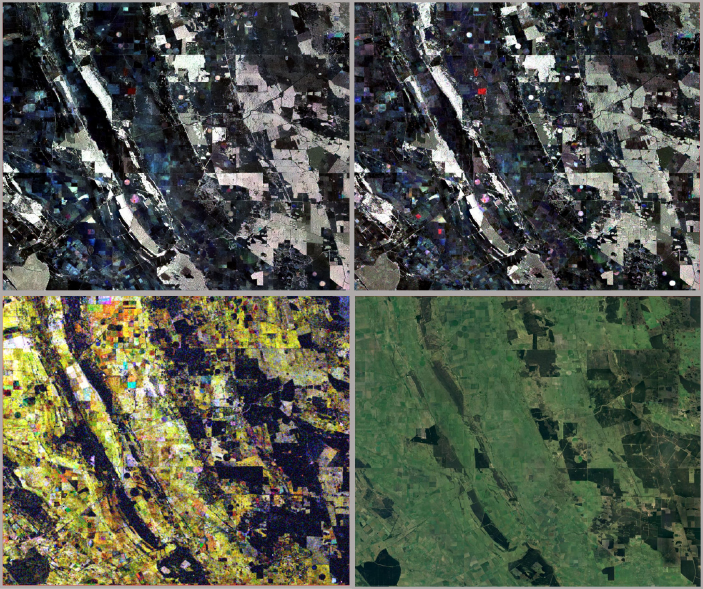}
    \caption{Contextual comparison of ROI with example image of each SAR data cube products and an RGB image. Top Left: VH, Top Right: VV, Bottom Left: CC, Bottom Right: Google satellite (Map data: Google, Imagery @2020 TerraMetrics).}
    \label{fig:VHVVCCNorm}
\end{figure}

Three Sentinel-1 SAR data cubes were used to develop SARGDV; a like-polarised data cube (VV), a cross-polarised data cube (VH), and an InSAR coherence data cube (CC). Each image of the data cubes had been processed to have a spatial resolution of $\approx{30}\text{m}$ (figure \ref{fig:VHVVCCNorm}) \citep{Castellazzi.et.al.2019}. For training and validation of the model, GDV data from the \hyperlink{http://www.bom.gov.au/water/groundwater/gde/map.shtml}{GDE Atlas} was used to classify the pixels pertaining to GDVs and non-GDVs of the data cube imagery.

\subsubsection{SAR data cubes}
Both the VV and VH data cubes are composed of 30 SAR images stacked in chronological sequence. Each image is a $2044\times1433$ raster of SAR intensity values. The images were acquired between 04/Jan/2017 to 11/Jan/2018, equating to one year of imagery with approximately 12 days between the capture of each image in sequence \citep{Castellazzi.et.al.2019}. Each pixel is thus associated with 30 values representing the SAR measurements of that location, as a time series.

The CC data cube is comprised of 29 images at the same resolution, stacked chronologically. Each pixel represents a correlation coefficient value ranging between 0 --- 1, computed as a measure of variance for each VV image pair, in sequence (see \citealt{Castellazzi.et.al.2019}).

The combination of these 3 data cubes equates to $8{,}787{,}156$ time series, each consisting of 89 pixels.

\subsubsection{Binary map from GDV data}
\begin{figure}[!h]
    \centering
    \includegraphics[scale = 0.65]{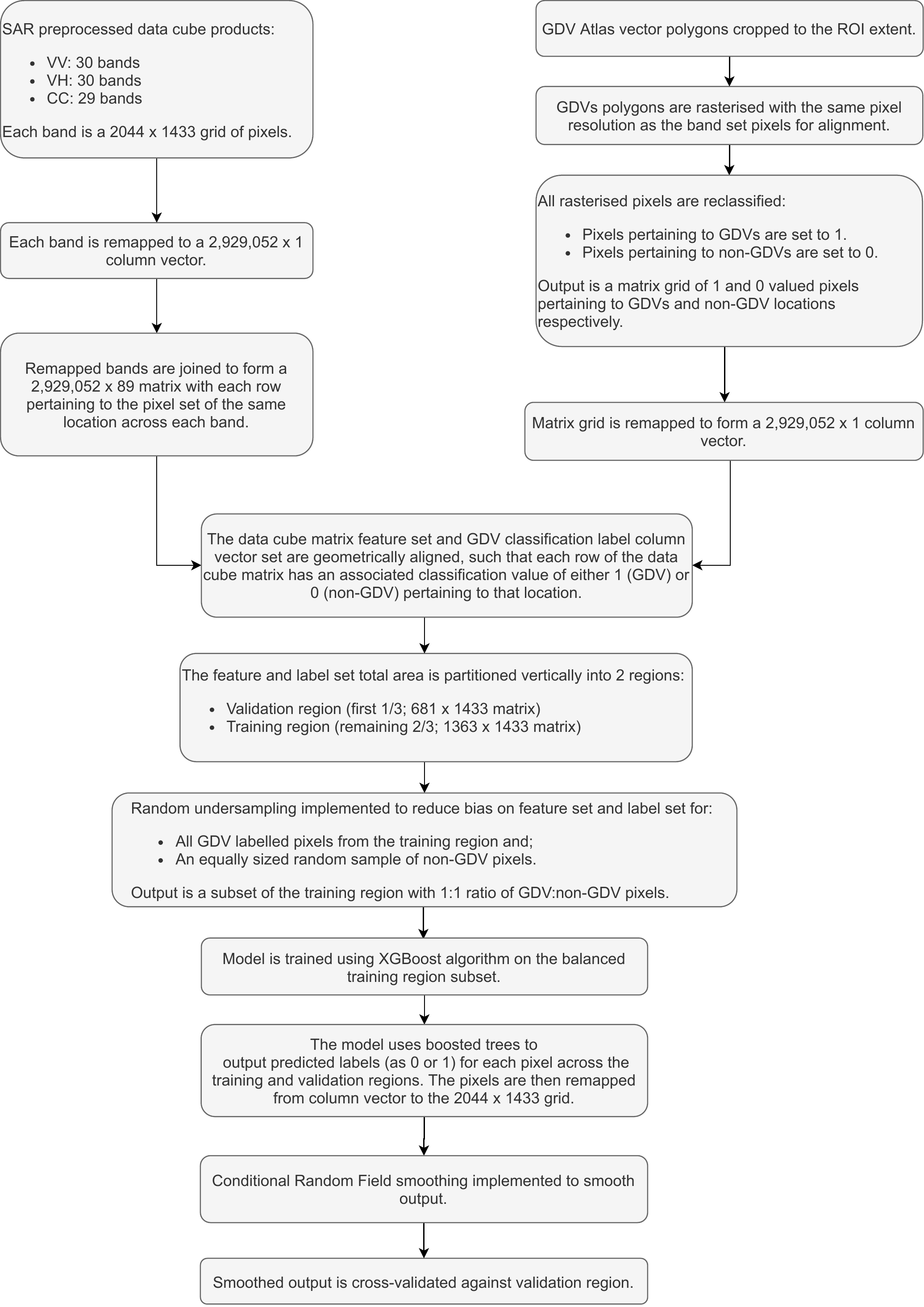}
    \caption{Conceptual diagram of methods workflow in sequence.}
    \label{fig:concept_diagram}
\end{figure}

For clarity, a conceptual diagram has been produced incorporating the proceeding sequence of key steps of the methods in a simplified manner (figure \ref{fig:concept_diagram}). The coordinate reference system used for processing was the WGS84 datum (\url{https://spatialreference.org/ref/epsg/wgs-84/}).

\begin{figure}[!h]
    \centering
    \includegraphics[scale = 2.4]{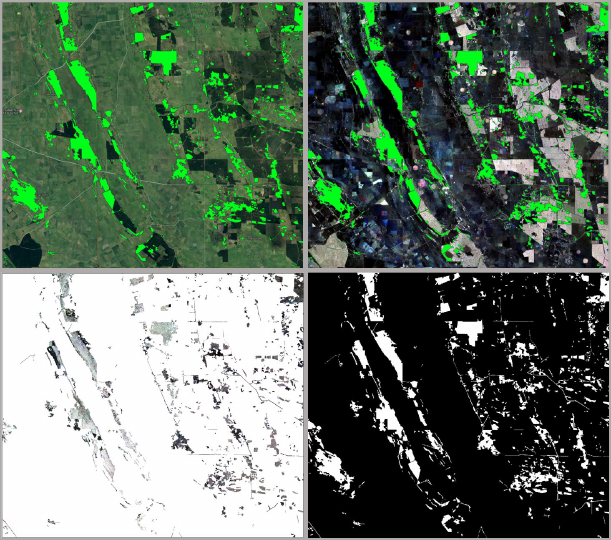}
    \caption{Rasterisation processing method illustrated in sequence. Top Left: Stage 1) GDV vectors overlaid on a satellite base map (Map data: Google, Imagery @2020 TerraMetrics), Top Right: Stage 2) GDV vectors overlaid on VV Band 1 SAR image base map for pixel extraction, via clipping tool with a threshold of $>=$ 50\% GDV vector overlap of the V pixels, Bottom Left: Stage 3) resulting raster of the extracted VV band 1 SAR pixels (GDV pixels coloured, non-GDV pixels set as white by default), Bottom Right: Stage 4) extracted pixels set to 1 (GDV) and all other pixels set to 0 (non-GDV), and coloured white and black respectively for visualisation.}
    \label{fig:Processing_method}
\end{figure}

The GDV data was obtained as vector polygons from the GDE Atlas data portal (\url{http://www.bom.gov.au/water/groundwater/gde/map.shtml}); a comprehensive inventory of potential GDEs across continental Australia \citep{SKM.GDE.Atlas.2012}. The data set provided a suitable set of samples for the given ROI (figure \ref{fig:Processing_method}).

The GDV vector data was cropped to the extent of the ROI, and subsequently used to crop the associated pixels of the VV SAR image. 
The values of these cropped pixels were changed to 1 (classifying as GDV), and all other pixel values in the raster were assigned to 0 (classifying as non-GDV), creating a binary raster (figure \ref{fig:Processing_method}).
The binary raster is used in the following model training stage to provide an associated classification (corresponding to GDV or non-GDV) for each pixel of each SAR data cube time series.

\subsubsection{Training and validation sets}
\begin{figure}[!h]
    \centering
    \includegraphics[scale = 2.0]{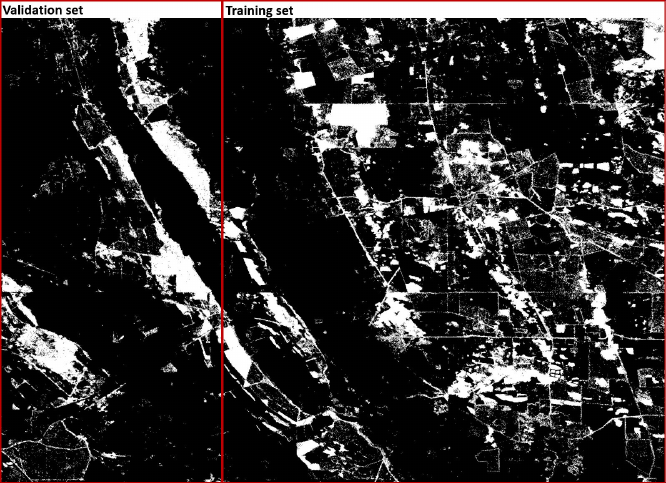}
    \caption{Unsmoothed model output of validation and training region predicted pixels within ROI.}
    \label{fig:Unsmoothed_res}
\end{figure}

The binary raster and each of the SAR data cube images are partitioned by vertically cropping the ROI extent ($2044\times1433$ pixels) into a training region and a validation region, comprising two thirds ($1363\times1433$ pixels) and one third ($681\times1433$ pixels) of the ROI, respectively (figure \ref{fig:Unsmoothed_res}).

The ratio of non-GDV to GDV class pixels in the ROI was approximately 9:1. Bias in boosted tree learners can be difficult to manage when data classes have a large imbalance, as upon training it is expected to heavily weight in favour of larger class (non-GDV). To resolve this class imbalance, the statistical method of random undersampling \citep{Xu-YingLiu.etal.2009} was implemented on the training region (see \url{https://machinelearningmastery.com/random-oversampling-and-undersampling-for-imbalanced-classification/}). Random undersampling was performed on all pixels classified as GDVs, and an equally sized random sample of non-GDV classified pixels from the training region. Thus, a class-balanced training set comprised of a 1:1 ratio of GDV:non-GDV pixels were extracted for training the model.

The binary raster is transformed from a raster matrix into a column vector where each row pertains to one pixel. Both the training, and validation raster matrices are transformed into row vectors, where each column pertains to one pixel. This enables matrix multiplication of the binary column vector with the associated row vector of the class-balanced training set, with each pixels from these different sets being correctly aligned in their ROI location. To ensure correct pixel location reference between the binary column vector, and the class-balanced training set, an additional vector was generated that reflects the location of pixels which were retained from the random undersampling. This is also included in the model. 

Lastly, though SARGDV is trained on the class-balanced set, the model is tested on all  pixels in the ROI. This allows us to evaluate areas where the GDE atlas agrees or disagrees with our model predictions across the entire ROI.

\subsubsection{XGBoost algorithm}
Extreme Gradient Boosting (XGBoost) is a reliable machine learning algorithm which uses boosted tree learners \citep{Chen.Guestrin.2016}.
The maximum tree depth $d$ and the number of boosting iterations $b$ were changed from their defaults to improve the accuracy of SARGDV. A large maximum tree depth of d = 50 was chosen to maximise usage of the available information, though this was expected to produce overfitting. To reduce overfitting, the maximum learning rate eta = 1 was applied. Performance on the test data suggests that overfitting was not critical. The $b$ was set to 40, it was identified that the convergence rate of the model slowed when approaching this amount. The specific parameters used for SARGDV can be found in the RSAR github repository (\url{https://github.com/frycast/rsar}). Using this algorithm in conjunction with the training set, and the hyper parameters specified, supervised learning was employed to output predicted GDVs at a pixel-by-pixel level. The resulting output is a column vector of GDV predicted pixels for each pixel of the ROI extent. The column vector is then transformed back to the original raster dimensions ($2044\times1433$ pixels) and is referred to as the unsmoothed output (figure \ref{fig:Unsmoothed_res}).

\subsubsection{Conditional Random Field smoothing}
\begin{figure}[!h]
    \centering
    \includegraphics[scale = 2.0]{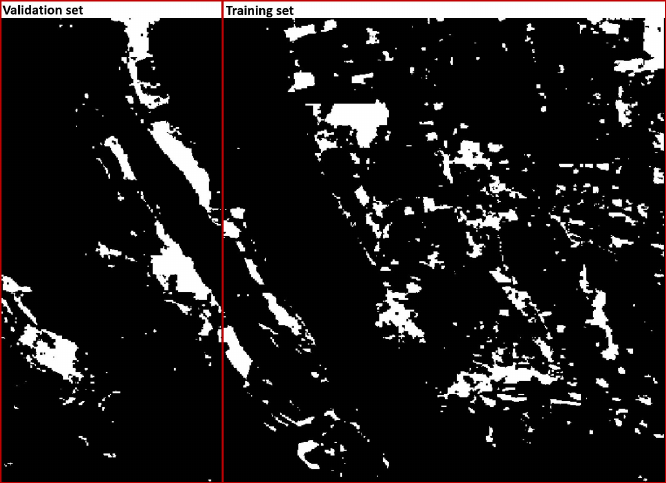}
    \caption{Smoothed model output of validation and training region predicted pixels within ROI.}
    \label{fig:Smoothed_res}
\end{figure}

To improve model performance, the CRF statistical method is applied to smooth the model output (see \citealt{Plath.etal.2009}). By implementing CRF smoothing, the pixels associated with predicted GDVs are retained if they have many neighbouring pixels that are also predicted GDVs, otherwise, they are discarded. This results in reduced prediction noise in the output (figures \ref{fig:Unsmoothed_res} and \ref{fig:Smoothed_res}), and is expected to improve the model performance by decreasing the false discovery rate of predicted GDVs.

\subsection{Model validation}

\subsubsection{Cross-validation with validation region}
To evaluate the performance of our model, a confusion matrix was generated assessing both the unsmoothed and smoothed outputs for the validation and training regions (table \ref{tab:conf_mat}). Additional performance metrics were collated to assess the accuracy of SARGDV (table \ref{tab:SARGDV2}).

\subsubsection{Benchmark against logistic regression}
Comparison of the XGBoost (non-linear classifier) and logistic regression (linear classifier) model performance demonstrates the non-linearity of the classification task, as the logistic regression classifier under-performed against the proposed XGBoost model (tables \ref{tab:SARGDV2} and \ref{tab:Log_reg}).

\subsubsection{ROC and PRC plots}
\begin{figure}[!h]
    \centering
    \includegraphics[scale = 3.63]{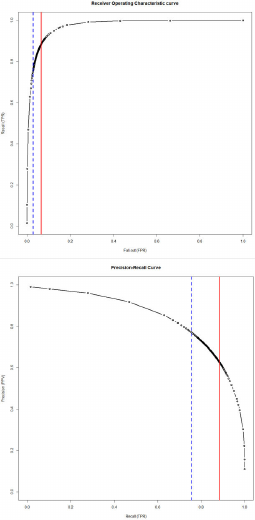}
    \caption{Top: ROC plot. Bottom: PRC plot. Threshold values: blue dashed line: p = 0.9 (model threshold), red solid line: p = 0.2 (benchmark threshold).}
    \label{fig:ROC-PRC}
\end{figure}

For further validation of the performance of SARGDV, two common classifier model validation measures: a ROC, and PRC plot (figure \ref{fig:ROC-PRC}) were generated with a chosen boundary threshold (p = 0.9), alongside an additional benchmark threshold (p = 0.2) for comparative evaluation. These plots offer a threshold-free measure for evaluation of SARGDV, providing an overview of the range of performance over variable thresholds rather than an evaluation that is tied to singular defined thresholds. 

The ROC plot illustrates the trade-off between the fallout (false positive rate) and the recall (true positive rate) of SARGDV, while the PRC plot provides validation of the reliability of SARGDV against false positives by examining the precision (the correct predictions among the positively predicted pixels), against the recall (see \citealt{Saito.Rehmsmeier.2015}).

\subsubsection{Depth to water interpolation over ROI}
\begin{figure}[!h]
    \centering
    \includegraphics[scale = 1.80]{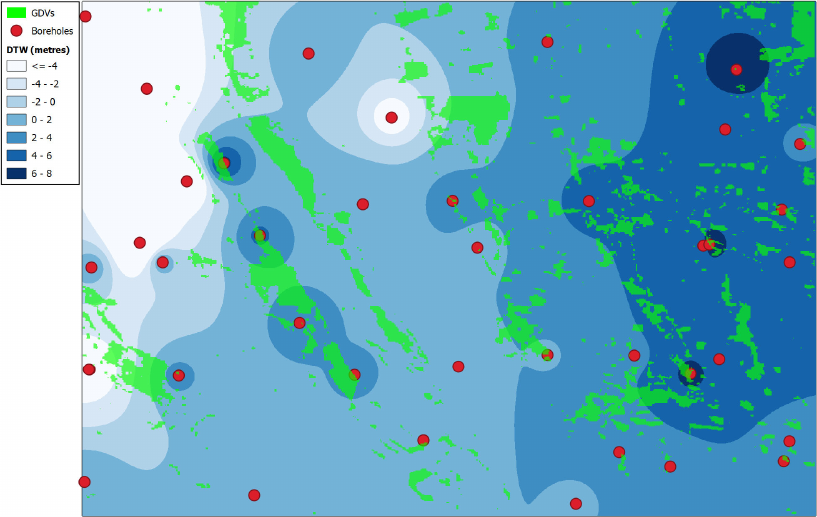}
    \caption{Inverse Distance Weighting interpolation map of DTW of 44 boreholes within the ROI with predicted GDVs. DTW varies seasonally (data not known).}
    \label{fig:IDW_interpolation}
\end{figure}

An inverse distance weighting (IDW) map was generated to interpolate the DTW across the ROI (see \citealt{johnston2004inverse}) (figure \ref{fig:IDW_interpolation}). To provide a more accurate impression of the expected groundwater levels within the ROI, the borehole data set was filtered. 
Measurements without DTW values were removed, and only observations collected from 2015 to the present day were used, as a period of approximately five years was assumed to offer a reasonable basis for evaluating the contemporary hydrological activity across the ROI. Of the remaining data set, only 2 samples had an observation date before 2019; these samples were removed to offer a more accurate view of the current DTW across the ROI. 
After filtering, 44 boreholes samples remained, offering a useful distribution within the ROI.

The range of observed DTW readings in the borehole data set was $-16.8$m to $7.68$m, with negative values indicating the water level being above ground. This DTW range offers a reasonable basis for validation of GDVs occurrence as vegetation in proximity to shallow groundwater (within 10m) offer higher likelihood of being GDVs \citep{Doody.etal.2019}.
The IDW map was processed with a distance coefficient of 2, the output resolution was set to $2044\times 1433$ (as with the data cube images), and the coordinate reference system was set to WGS84 (see https://spatialreference.org/ref/epsg/wgs-84/).

\section{Results}

The following section uses terminology relating to metrics used in the evaluation of classification models. These metrics have been defined in the tables section of this paper (table \ref{tab:SARGDV2}).
The SARGDV model showed excellent results with respect to its FDR, TPR, TNR and FOR, for both the smoothed (figure \ref{fig:Smoothed_res}) and unsmoothed (figures \ref{fig:Unsmoothed_res}) outputs, over the validation set. 
Both the unsmoothed and smoothed output, from SARGDV are addressed to emphasize the degree of performance that is contributed by the use of CRF.

The SARGDV model performed well over the study region. 
For the chosen model threshold ($p$ = 0.9), the ROC plot indicated a TPR of 76\% with an FPR of 3\%, while the PRC plot returned a precision of 77\% demonstrating that the model performance was well balanced for the chosen threshold (figure \ref{fig:ROC-PRC}). In contrast, the ROC plot for the chosen benchmark threshold ($p$ = 0.2) showed that an increase in the TPR to 89\% was met with the trade-off of an increase in the FPR to 7\%, while the PRC plot indicated a sizeable decrease in the resulting precision to 63\%.

Comparison of SARGDVs results (table \ref{tab:SARGDV2}) against the chosen logistic regression model benchmark results (table \ref{tab:Log_reg}) demonstrates an overall better predictive performance is achieved from the use of the boosted tree learner. 

Comparison of the validation set shows the smoothed output provides a good balance of GDV predictions, with some minor trade offs.
Where there was only 1\% difference in the FDR between the logistic regression output and the unsmoothed output, the smoothed output achieved approximately 20\% less false discoveries than both the aforementioned outputs.
This was associated with approximately 12.7\% reduction in the TPR of the smoothed output, as compared with the unsmoothed output, which had the highest TPR, however this trade off was considered worthwhile as it increases the reliability of the model predictions, rather than trying to identify every GDV pixel and potentially acquiring a sizeable number of false positives. 
Additionally, the smoothed output also achieved approximately 3\% improvement over both the unsmoothed and logistic regression outputs.
The FOR was comparable across all three models.

\section{Discussion}
As GDVs are vegetation areas that access groundwater at some time in their life history, regional water resource management requires knowledge of GDV location and where possible, their ecological water demand. However, regional GDV maps and their relative ecohydrological attributes are often not available, outdated or limited \citep{Barron.et.al.2014}. 

Many GDV studies are constrained to local-scale evaluation, using expensive in-situ measurements and pertain to a singular species or ecotype \citep{Barron.et.al.2014, O'Grady.2010, Eamus.et.al.2006}. Additionally, many areas globally, are not aware of locations of GDVs and so economical broad-scale methods such as the SAR method reported are required to pinpoint areas, where vegetation may access groundwater. Once location is known, further studies can target groundwater requirements, especially for protected native species and then plan for management of groundwater levels in the face of potential current and future groundwater extraction and climate change.

While the challenge of producing regional GDV maps by scaling up from local-scale studies is not a new idea \citep{Froend.Sommer.2010}, it has not yet been satisfactorily concluded via an efficient methodology. To date, there have been numerous attempts to meet this requirement using remotely sensed data sets \citep{Barron.et.al.2014, Werstak.Maus.2012}. As demonstrated from the method presented in this paper, the SARGDV model offers efficient, regional GDV mapping.

\subsection{Generalisability of the model}

Improved generalisability is expected to offer improved transferability of the model across regions both within Australia and internationally. Thus, several development choices were made to enable greater generalisability of the model.

While the SARGDE model produced in \cite{Castellazzi.et.al.2019} is comprised of a single equation, where the thresholds for classifying GDVs was heavily influenced by assumptions of expected GDE responsiveness, SARGDV does not use specified assumptions and instead learns the important feature weightings from the supplied data to develop optimal thresholds for the classification task. This provides the opportunity for more of the vegetation characteristics information to be captured during the model training stage, and utilised to improve model performance. The XGBoost algorithm was selected for SARGDV as it is considered an efficient, scalable, state-of-the-art machine learning algorithm that optimises well for general classification tasks \citep{Chen.Guestrin.2016}.

Additionally, CRF smoothing encourages generality by reducing the number of false positives, in a tradeoff with slightly increased false negatives. The intention was to create a more trustworthy model, having a lower FDR, that also translates into a bias reduction.

Further research is required to evaluate the SARGDV models ability to generalise to new SAR data and offer useful transferability to new regions. Should the model perform well, this method may provide numerous global institutions, hydrological management personnel, and government bodies, with a low cost, reproducible method to map GDVs and further support the preservation of these significant ecological communities.

\subsection{Model performance comparison}
Comparison of SARGDV against similar models highlights key areas of contrast. The groundwater-dependent ecosystem mapping (GEM) method produced by \cite{Barron.et.al.2014} was derived from a two-step classification approach. Initially the normalised difference vegetation index (NDVI) and also the normalised difference water index (NDWI) were calculated for two Landsat thematic mapper images, each acquired at the end of the wet and dry seasonal periods respectively, before using the ISODATA unsupervised classification technique to find land cover clusters. This was followed by an analysis of the identified class statistical centroids, with a focus on separating the potential GDE-related classes from other land cover classes. The GEM model was able to achieve 91\% similarity with field results but with a bias toward better results for “regions with a distinct and prolonged dry period”. Thus, generalisability of the model was sacrificed for a more focused tuning of the model toward these specific regions.

The SARGDE model is derived as an index equation combining three statistical observations, which have been normalised across each scene, to discriminate between GDEs and non-GDEs. The observations are formed from domain specific assumptions of vegetation characteristics, which are purported to indicate high likelihood of GDE occurrence, and these are used as the parameters for the index equation. The index returns a high value when the likelihood of GDE occurrence is high, however as it is the first iteration of the models design, the parameters are all equally weighted. No optimisation studies have been conducted yet to identify how the index performance may improve from more optimal weighting.

While SARGDV uses input data with the same processing method, and both models predict GDE occurrence, SARGDE and SARGDV are different in their model design. SARGDV does not use an index equation with pre-assigned parameters, instead its discriminant function is formed with parameters that have been optimised through model training using the XGBoost boosted algorithm. The SARGDV model learns from the input data set, the combination of pixels flagged as GDV, and the 89 SAR data product images. Prior to the output there is also the addition of a smoothing procedure that takes place to provide more conservative predictions and improve model performance.

SARGDE achieved 91\% identicality when compared to the GDE Atlas data for a different subset of the Mount Gambier region to the reported study region. For comparison, the accuracy from SARGDV on the subset of Mount Gambier region used in this paper were 96\% and 94\% for the smoothed and unsmoothed outputs, respectively, where accuracy is (TP + TN)/sample size. While the comparison is not over the same study location of Mount Gambier, these results indicate that the SARGDV model shows better performance over a similar local region, which is expected to have the same or similar set of GDV species.

Visual analysis of the SARGDV outputs suggest that the unsmoothed output may be picking up GDVs that have not been included in the GDE Atlas data set (figure \ref{fig:Unsmoothed_res}). Field-based surveys within the ROI would prove useful for obtaining ground-truth observations to validate these results and determine which of these outputs are better suited for general use.

\subsection{Borehole validation} 
Evaluation of the IDW map outlined that many of the predicted GDVs are located over areas of shallow groundwater (figure \ref{fig:IDW_interpolation}). This offers additional support to the validity of the predicted GDVs as shallow groundwater is associated with higher likelihood of GDV presence \citep{Doody.etal.2019}. 
However, given the limited number of boreholes and seasonal variation of DTW, the IDW map can only offer a suggestive interpolation across the ROI for the actual DTW, and therefore the level of certainty for validation via this method is also constrained by this.

\subsection{Implications for environmental management}
The study region is well known for its significant plantation wood production. Detailed studies by \cite{benyonImpactsTreePlantations2006} and \cite{doody.2011.direct} have clearly identified that plantations around the region access groundwater, extracting up to 670 mm year-1 depending on the depth to water-table over which they are located. The region is also known for its shallow karst groundwater system \citep{Emmett.Telfer.1994, Grimes.1994} and extensive wetland network. A desktop analysis by \cite{harding.2012.delivering} indicates that the south-east region of South Australia, contains the largest area of potential groundwater-dependent ecosystems in the state, with over a third located in this region. Hence, unmanaged extraction of groundwater resources by plantation forests in the region is likely to impact these important non-plantation wetlands and their associated GDVs by disrupting the hydrological processes which support them. To prevent wetland and GDV decline in future, areas may be targeted as \say{non-plantation} zones, based on mean depth to groundwater. \cite{benyonImpactsTreePlantations2006} demonstrated that plantation groundwater extraction was not species dependent, but rather depth to water-table dependent and once water-table depth exceeded 6m, trees were less likely to extract substantial amounts of groundwater.

Development of the method within, provides a novel, rapid application, economical tool to identify GDVs. Providing the method is transferable to broader regions and climates (as demonstrated by \citealt{Castellazzi.et.al.2019}), this tool will allow for significant advances in how GDVs are identified, worldwide. This is especially important when water extraction is related to industries such as mining and agriculture.  Currently, mining companies in Australia, struggle to adequately identify and map groundwater-dependent ecosystems in their environmental impact assessments, leading to significant delays in government assessments. Additionally, if GDVs have been identified, there are few rapid, cost-effective means by which to monitor the GDVs as mining or irrigated agricultural extraction proceeds (for example). In Australia, there are very strict rules related to coal seam gas and large coal mining and groundwater-dependent ecosystems. All groundwater-dependent ecosystems (including GDVs) must be identified and monitored during the course of mining to mitigate risks of environmental impacts \citep{Doody.etal.2019}.

\subsection{Further research direction}
While the proposed model performs very well in the study region, limitations exist that will inform further research. Both the limited training data, and validation of the model over the sole ROI does not offer a suitable verification of the models capacity to generalise well to regions outside the ROI. Given more training data over several different regions, it is expected that SARGDVs capacity to generalise well to new regions will improve. The degree and variety of data that would be required to meet a suitable level of generalisability is not yet known. Testing of the generalisability and transferability of the model is essential to verifying that the model supports the development focus. Although it is not yet tested, we anticipate the model will perform well across different regions. Follow-up research will investigate this hypothesis further.

Additionally, there is merit in exploring the potential for improvements to model performance by tuning the model parameters to different values based on the distinct regions GDVs. This would include analytical works to identify specific features of GDVs that are shared or contrast across regions, in order to weight the respective features accordingly for improved classification. While this region specific tuning of the model would likely sacrifice generalisability, it may also serve to provide a basis to enhance model performance greatly, by enabling the classification of otherwise unidentified GDVs that may be missed by the current model. Research has not yet been conducted for model explanation, that is, how the model decides which input properties cause the GDV predictions, which may be achieved through model evaluation methods such as Shapley feature attribution \citep{fryer.2020}.

A substantial knowledge gap exists, related to when GDVs connect and disconnect with groundwater. It is highly unlikely that permanent connection is a common case for most GDVs, as groundwater levels flux due to seasonal precipitation inputs and extraction demand, especially during summer. The 12-day satellite return period of Sentinel-1 SAR opens the door to investigate temporal and seasonal changes in GDVs over the last 5--6 years of image capture. Temporal mapping, coupled with groundwater fluctuation data, can begin to provide insights into critical periods of time in groundwater extraction areas, for when groundwater pumping should be managed to ensure groundwater disconnection is prevented as much as possible or at least minimised. It is likely that with enough temporal GDV and groundwater depth data, thresholds of depth to groundwater can be identified, which lead to vegetation disconnection. Once such thresholds are elucidated, groundwater extraction can be managed to prevent decline in GDVs. In data poor regions, where groundwater data collection is not commonplace, advances in remote sensing of soil moisture will be required to help inform hydrological management of GDVs.
\section{Conclusion}

The threat of death and desiccation of GDVs has been reported at an international scale. GDVs offer several natural support mechanisms for both animal and human communities. However, continued decline of GDVs impacts the biodiversity and survival of many groundwater species of flora and fauna. Over exploitation of groundwater through poor water management and industry use are significant factors toward impacting the health and viability of GDVs. Preservation of these ecosystems thus relies on efficient and cost-effective methods to identify GDVs at scale, which can support more informed and sustainable regional water resource management. The SARGDV method that we present offers a reproducible, economical, and accurate means to identify GDVs at scale, and may further enable strategic preservation of GDVs long term. Our model has demonstrated high accuracy for predicting the presence of GDVs in the study ROI, showed improved performance in all metrics including FDR and TPR when compared against the logistic regression benchmark, and performed well against other similar GDV prediction models. As SARGDV has been designed for generalisability, it is expected that the model produced from this research will perform better with additional SAR data cube products for other regions worldwide. The method outlined also offers the potential for transferability to other regions of variable climates, within continental Australia, and around the world.

Ensuring the long term protection of GDVs is of vital importance in the effort to prevent unsustainable loss of biodiversity and severe impact to the communities that rely on these natural ecosystems. Due to its feasibility and ease of reproducibility, our method may offer institutions an efficient means to strategically manage the preservation of their GDV inventories through improved water regional resource management, and designating protected vegetation zones to avoid impact from industry. This research sets the foundation for further research that aim to develop the model toward improved generalisation to regions outside of the study ROI. It is expected that, with the addition of more data to train the SARGDV model, further latent vegetation characteristic information can be utilised that may improve overall model performance for predicting true positive GDVs with better accuracy. Exposure of the model to training data that contains a larger variety of GDV and non-GDV species is likely to improve performance in detecting true GDVs, and offset potential false positives.

\section{Acknowledgements}

The authors would like to thank the European Space Agency’s Copernicus Program for Sentinel-1 IW data in the SLC format 
(\href{https://scihub.copernicus.eu/}{scihub.copernicus.eu})
and the Alaska Satellite Facility for providing a user-friendly portal to navigate and acquire most types of open SAR data used in the research (\href{https://search.asf.alaska.edu/}{search.asf.alaska.edu}). 
We would like to thank the Bureau of Meteorology for providing the GDE Atlas data set available in their online portal (\href{http://www.bom.gov.au/water/groundwater/gde/map.shtml}{bom.gov.au/water/groundwater/gde/map.shtml}), 
the government of SA for their landcover data set available through their online portal 
(\href{https://data.sa.gov.au/}{data.sa.gov.au}), 
and WaterConnect for the borehole data set (\href{https://www.waterconnect.sa.gov.au/Systems/GD/Pages/Default.aspx}{waterconnect.sa.gov.au/Systems/GD/Pages/Default.aspx}). Hien Nguyen and Mason Terrett are funded by Australian Research Council grant DE170101134.

\bibliography{z_references}
\bibliographystyle{apacite}

\begin{table}[!h]
\caption{SARGDV confusion matrix results for the training and validation set.}
\begin{tabular}{c|c|c|c|c|c|c|c}
      & Function               & TP    & TN    & FP    & FN  \\ \hline
  Validation   & unsmoothed    & 89939 & 829104 & 39536 & 17294 \\ 
  set          & smoothed      & 76493 & 859613 & 9027 & 30740 \\ 
  \hline
 Training       & unsmoothed    & 192687 & 1605882 & 154610 & 0 \\ 
 set            & smoothed      & 152761 & 1723873 & 36619 & 39926 \\\hline
    \label{tab:conf_mat}
\end{tabular}
\end{table}

\begin{table}[!h]
\caption{SARGDV confusion matrix performance metrics for the training and validation set.
False Discovery Rate (FDR): the proportion of samples classified as positive that are actually negative;
True Positives Rate (TPR): the proportion of samples classified as positive that are actually positive; and 
True Negative Rate (TNR), the proportion of samples classified as negative that are actually negative.
False Omission Rate (FOR): the proportion of samples classified as negative that are actually positive. 
}
\begin{tabular}{c|c|c|c|c|c|c|c}
       & Function               & FDR   & TPR   & TNR   &   FOR     \\ \hline
   Validation   & unsmoothed    & 0.305 & 0.839 & 0.955 &   0.020   \\ 
   set          & smoothed      & 0.106 & 0.713 & 0.990 &   0.035    \\ 
   \hline
 Training       & unsmoothed    & 0.445 & 1     & 0.912 &   0       \\ 
 set            & smoothed      & 0.193 & 0.793 & 0.979 &   0.023    \\\hline
    \label{tab:SARGDV2}
\end{tabular}
\end{table}

\begin{table}[!h]
\caption{Logistic regression model performance metric results for the training and validation set.}
\begin{tabular}{c|c|c|c|c|c|c}
                  & FDR    & TPR    & TNR       & FOR   \\ \hline
   Validation set & 0.318 & 0.737   & 0.958     & 0.033 \\ 
   \hline 
  Training set    & 0.604 & 0.754   & 0.874     & 0.030 \\ \hline
  \label{tab:Log_reg}
\end{tabular}
\end{table}

\begin{figure}[!h]
    \centering
    \includegraphics[scale = 1.65]{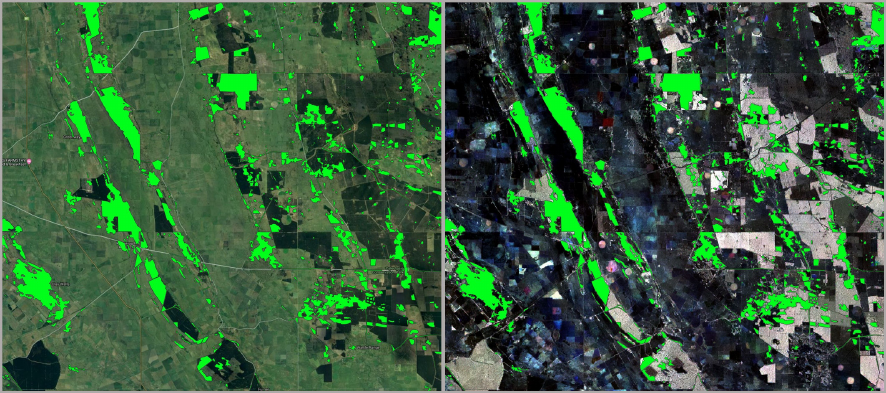}
    \caption{GDV vectors from GDE Atlas within the ROI, overlaid on both a satellite and SAR image base maps. Left: Google Satellite (Map data: Google, Imagery @2020 TerraMetrics), Right: VV band 1 SAR image.}
    \label{fig:GDV_VV_atlas}
\end{figure}

\end{document}